\documentclass[aps,prd,superscriptaddress,preprintnumbers,nofootinbib,12pt]{revtex4-2}

\usepackage{graphicx,amsmath,amssymb}
\usepackage{bigints}
\usepackage{hyperref}
\usepackage{bm}
\usepackage{ulem}
\usepackage{hyphenat}
\usepackage{lineno}

\hypersetup{colorlinks=true, linkcolor = [rgb]{0,0.08,0.45}, citecolor = [rgb]{0,0.08,0.45}, urlcolor = [rgb]{0,0.08,0.45}}

\newcommand{\req}[1]{(\ref{#1})}
\newcommand{\lb}{\label}
\newcommand{\nn}{\nonumber}

\begin{document}


\title{Asymptotic gauge symmetry and UV extension of the nonperturbative coupling in holographic QCD}


\author{Guy~F.~de~T\'eramond}
\email[]{guy.deteramond@ucr.ac.cr}
\affiliation{Laboratorio de F\'isica Te\'orica y Computacional, Universidad de Costa Rica, 11501 San Jos\'e, Costa Rica}

\author{Arpon~Paul}
\email[]{paul1228@umn.edu}
\affiliation{School of Physics and Astronomy, University of Minnesota, Minneapolis, Minnesota 55455, USA}

\author{Hans~G\"unter~Dosch}
\email{h.g.dosch@gmail.com}
\affiliation{Institut f\"ur Theoretische Physik der Universit\"at, D-69120 Heidelberg, Germany}

\author{Stanley~J.~Brodsky}
\email[]{sjbth@slac.stanford.edu}
\affiliation{SLAC National Accelerator Laboratory, Stanford University, Stanford, California 94309, USA}

\author{Alexandre~Deur}
\email[]{deurpam@jlab.org}
\affiliation{Thomas Jefferson National Accelerator Facility Newport News, Virginia 23606, USA}

\author{Tianbo~Liu}
\email{liutb@sdu.edu.cn}
\affiliation{Key Laboratory of Particle Physics and Particle Irradiation (MOE), Institute of Frontier and Interdisciplinary Science, Shandong University, Qingdao, Shandong 266237, China}
\affiliation{Southern Center for Nuclear-Science Theory (SCNT), Institute of Modern Physics, Chinese Academy of Sciences, Huizhou 516000, China}

\author{Raza~Sabbir~Sufian}
\email[]{gluon2025@gmail.com}
\affiliation{Department of Physics, New Mexico State University, Las Cruces, New Mexico 88003, USA}
\affiliation{RIKEN-BNL Research Center, Brookhaven National Laboratory, Upton, New York 11973, USA}
\affiliation{Physics Department, Brookhaven National Laboratory, Upton, New York 11973, USA}

\date{\today}



\begin{abstract}

\vspace{-4pt}

\begin{center}
(HLFHS Collaboration) \\
\vspace{10pt}
\includegraphics[scale=0.18]{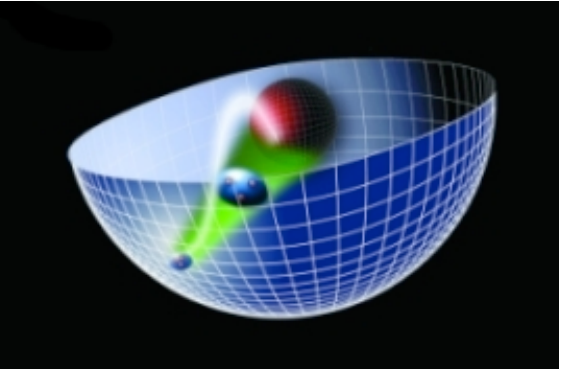}\\
\end{center}

\vspace{-8pt}

We extend our recent analytic study of the strong coupling $\alpha_{\rm eff}$ in the nonperturbative and near-perturbative regimes~\cite{deTeramond:2024ikl} by imposing rigorous renormalization-group results from asymptotically free gauge theories at $Q^2 \to \infty$. The asymptotic boundary conditions modify the scaling properties of $\alpha_{\rm eff}$  at large values of the momentum transfer $Q^2$, and lead to a scale-dependent confinement strength $\kappa(Q^2)$. This requires that both $\kappa(Q^2)$ and $\alpha_{\rm eff}\left(Q^2, \kappa(Q^2)\right)$ remain holomorphic in the complex $Q^2$ plane, except at the physical cuts associated with the heavy-quark thresholds and the singularity flow trajectory studied in~\cite{deTeramond:2024ikl}. For color $SU(3)$, a precise connection is found between the scaling exponent of $\kappa(Q^2)$ in the ultraviolet,  the value of the infrared fixed point of the strong coupling, and the number of flavors in agreement with observations. The nonperturbative analytic model gives an accurate description of the strong coupling at all scales, up to the highest available data.

\end{abstract}

\maketitle

\tableofcontents

\newpage

\section{Introduction}

For strongly coupled quantum field theories, the AdS/CFT correspondence~\cite{Maldacena:1997re, Gubser:1998bc, Witten:1998qj}, the duality between gravity in anti-de Sitter space (AdS) and conformal field theory (CFT), provides a classical gravity approximation, which is often mathematically tractable, thus an ideal tool to explore the nonperturbative content of quantum chromodynamics (QCD). In particular, the strong QCD coupling $\alpha_s$ at large distances, which is required to describe the basic constituents of  hadrons and nuclei,  is not amenable to the usual perturbative iteration at short distances, and holographic and other nonperturbative methods have been proposed to unravel its complex nature in the low-energy infrared (IR) domain of QCD~\cite{Deur:2016tte}.

The AdS/CFT duality~\cite{Maldacena:1997re} originates in string theory and is the most successful concrete realization of the holographic principle~\cite{Bekenstein:1973ur, Hawking:1975vcx,  tHooft:1993dmi, Susskind:1994vu}. However, QCD as a fundamental physical theory based on color $SU(3)$ is not supersymmetric, neither conformal nor large $N_C$.  In practice, one is thus led to construct gravity duals of QCD which incorporate confinement and basic QCD properties in physical spacetime, a bottom-up approach known as AdS/QCD, or holographic QCD~\cite{Gross:2022hyw}.  In particular,  the holographic light-front approach to QCD (HLFQCD) is based on Dirac’s Hamiltonian front form of relativistic dynamics~\cite{Dirac:1949cp}. Light-front (LF) quantization is frame-independent, a property that enables a precise mapping of the equations of motion in a higher-dimensional gravity theory to semiclassical Hamiltonian equations in Minkowski spacetime at the boundary of AdS space~\cite{deTeramond:2008ht}. This approach leads to an effective computational framework to describe the bound states of the fundamental constituents in the boundary quantum field theory~\cite{Brodsky:2014yha}.

Within the light-front holographic framework, a new procedure was devised in Ref.~\cite{Brodsky:2010ur} to map the effective coupling $g_s(z)$ in the AdS space to the physical coupling $g_s(Q^2)$ in the physical spacetime. It accounts for the measurements of $\alpha_s$ in the IR domain~\cite{Deur:2016tte} without introducing arbitrary parameters.  The variable $z$, the fifth dimension of the AdS space, is also known as the holographic variable and has an exact mapping to the invariant transverse distance between the hadron constituents in the light front~\cite{Brodsky:2010ur}.

The analytic properties of quantum field theory require the continuity of physical observables to describe the transition between the IR and ultraviolet (UV) domains. We have therefore chosen in~\cite{deTeramond:2024ikl}, as well as in~\cite{Brodsky:2010ur}, the definition of $\alpha_s$ as an effective charge, which is an observable~\cite{Grunberg:1980ja, Grunberg:1982fw}. This allows us to extend the nonperturbative description of the strong coupling in HLFQCD into the transition, near-perturbative, domain using analytic continuation methods.  Effective charges extend the running coupling definition to include long-range dynamics in addition to short-range processes, namely iterative quantum loops~\cite{Deur:2016tte, Deur:2023dzc}. This scheme restores the scheme-dependent running coupling in the renormalization group equation (RGE) to the status of an observable, as required by the analytical construction of our approach.

In this article, we describe the extension of the effective strong coupling $\alpha_{\rm eff}$ introduced in~\cite{deTeramond:2024ikl} to arbitrarily large $Q^2$, a domain in which the symmetry group structure of asymptotically free gauge theories, determine the evolution of the system. In previous holographic studies of heavy quark spectroscopy~\cite{Gutsche:2012ez, Dosch:2016zdv, Nielsen:2018ytt}, a scale dependence of the confinement strength $\kappa$ was introduced following the results of the heavy quark effective theory (HQET)~\cite{Shuryak:1981fza, Isgur:1991wq}. In the present framework, this dependence on the effective scale corresponds to the presence of heavy-quark flavors and their influence at large $Q^2$. Following this assumption, we examine the flow of $\kappa(Q^2)$ as $Q^2 \to \infty$, consistent with the asymptotic boundary conditions and the analytic structure found in~\cite{deTeramond:2024ikl}.  The effective charge at different scales has also been computed with the Dyson-Schwinger and lattice methods in~\cite{Binosi:2016nme, Cui:2019dwv}.

The contents of this article are as follows. Following~\cite{Brodsky:2010ur}, we briefly review in Sect.~\ref{IR} the holographic approach to hadronic physics and describe the derivation of the strong coupling in the deep IR domain by mapping the effective coupling in the AdS action to the light front. We also discuss the specific connection between the coupling $g_s(z)$, which encodes the IR dynamics with the confinement potential in AdS. In Sect.~\ref{IRUV} we discuss the salient points of our recent paper~\cite{deTeramond:2024ikl}, which describes the transition regime between the IR and UV domains based on analytic continuation methods. Then,  in Sect.~\ref{LQ2} we examine the large-$Q^2$ behavior of the effective coupling in the context of~\cite{deTeramond:2024ikl}.  These three sections set the stage for the extension of the model to the UV domain. Guided by the scale dependence of the confinement strength $\kappa$ from heavy-quark spectroscopy,  a scaling procedure is introduced in Sect.~\ref{fu} by studying the flow of $\kappa(Q^2)$. It allows us in Sect.~\ref{UV} to impose rigorous renormalization group results at asymptotic infinity, consistent with the analytic structure of the model. It leads to precise relations between the scaling behavior with the number of open-quark flavors in the deep UV and the value of the IR fixed point, independently of any dimensionful parameter. We discuss in Sect.~\ref{exp}, the dynamics of thresholds from heavy quarks, and derive UV sum rules that bound the spread of possible solutions compatible with deep UV constraints. We compare in Sect.~\ref{allscal} our model prediction for the effective strong coupling with the available data at all scales, thereby including the region where finite-order perturbation theory becomes more and more reliable as the virtuality is increased towards the multi-TeV domain.  Conclusions are given in Sect.~\ref{CaO}. Some technical aspects are described in the appendices.

 \section{Strong coupling in the IR domain and holographic light-front QCD\lb{IR}}

In holographic QCD a hadron bound state is described by a classical field propagating in AdS space. Anti-de Sitter AdS$_{d+1}$ is the maximally symmetric $d+1$ space with negative constant curvature and a $d$-dimensional flat space boundary, Minkowski spacetime with coordinates $x^\mu$. The line element is
\begin{align} \label{AdSm}
ds^2  &= g_{MN}dx^M dx^N  \nn \\
&= \frac{R^2}{z^2} \left(\eta_{\mu \nu} dx^\mu dx^\nu - dz^2\right),
\end{align}
where $R$ is the AdS radius and $z$ is the holographic variable. The asymptotic boundary of AdS space is defined at $z = 0$.

An arbitrary integer-spin $J$ hadron is described by a bilinear action in AdS$_{d+1}$  for a tensor field $\Phi_J = \Phi_{N_1 \dots N_J}$ in the presence of a dilaton profile $\varphi(z)$ responsible for the confinement dynamics~\cite{deTeramond:2008ht, deTeramond:2013it}
\begin{align} \lb{SAdS}
  S = \int d^dx \,  dz \sqrt{g} \, e^{\varphi(z)}  \left(D_M \Phi_J D^M \Phi_J  - \mu^2 \Phi_J^2\right),
  \end{align}
where a contraction over all indices in the tensor fields is to be understood. The determinant of the metric tensor $g_{MN}$ is $g$,  $\mu$ is the AdS mass, and $D_M$ is the covariant derivative that includes the affine connection.   An important element for our purposes is the definition of the effective coupling $g_s(z)$ in the semiclassical gravity approximation~\cite{Aharony:1999ti, Gursoy:2007er, Pirner:2009gr, Brodsky:2010ur}
\begin{align}
e^{\varphi(z)} = \frac{1}{g_s^2(z)},
\end{align}
since it accounts for both the hadronic structure and the strong coupling in the IR domain.

In order to make contact with a physical hadron with four-dimensional momentum $P$  and invariant mass $ P_\mu P^\mu = M^2$, we write the AdS wave function as the product of a $z$-dependent wave function $\Phi(z)$ and a plane wave, $\Phi(x,z) \sim \Phi(z) e^{i P \cdot x}$, propagating in physical space-time. The variation of the AdS action leads to the wave equation
  \begin{align} \lb{AdSWEJ}
      \left[
   -  \frac{ z^{d-1- 2J}}{e^{\varphi(z)}}   \partial_z \Big(\frac{e^{\varphi(z)}}{z^{d-1-2J}} \partial_z   \Big)
  +  \frac{(\mu\,R )^2}{z^2}  \right]  \Phi(z) = M^2 \Phi(z), 
  \end{align}
after a redefinition of the AdS mass $\mu$, along with kinematical constraints that eliminate lower-spin degrees of freedom from the spectrum~\cite{deTeramond:2013it}.

We can express the AdS wave equation~\req{AdSWEJ} in its Schr{\"o}dinger form after the full separation of kinematics and dynamics~\cite{deTeramond:2008ht, deTeramond:2013it}. Using the substitution $\Phi_J(z) = z^{(d-1)/2 -J} e^{- \varphi(z)/2} \, \phi(z)$ we find 
\begin{align} \lb{AdSWE}
\left(-\frac{d^2}{dz^2} 
- \frac{1 - 4 \nu^2}{4 z^2}+ U(z) \right)  \phi(z) = M^2 \phi(z),
\end{align}
where  $\nu^2 = (\mu R)^2 + (d/2-J)^2$. The effective interaction potential $U(z)$ can be written in terms of the gravitational coupling $g_{s}$~\cite{deTeramond:2023qbo}
\begin{align} \lb{Ugz}
U(z) =  g_{s}(z) z^{-\gamma - 1}  \partial_z \left(z^{\gamma +1} \, \partial_z g_{s}^{-1}(z) \right),
\end{align}
with $\gamma = 2J - d$. This expression gives an explicit and simple connection between the gravitational coupling in AdS, $g_s(z)$, and the effective potential $U(z)$, which has a wide range of applications for the hadron spectrum.  We notice that the potential $U$ is not modified by rescaling the coupling $g_s(z)$ by a constant  $ \lambda$:  $g_s(z) \to \lambda g_s(z)$.

\subsection{Mapping to light-front physics and the physical strong coupling}

Starting from the QCD Lagrangian, one can express the Hamiltonian operator in terms of the dynamic quark and gluon fields~\cite{Brodsky:1997de}, and compute, in principle, the mass spectrum. However, its actual analytic computation is a task that has proven insurmountable in the context of physical QCD. Important insights follow from the profound  holographic connection between hadronic modes in the AdS space and the semiclassical approximation to QCD using LF quantization~\cite{Dirac:1949cp}.

In the chiral limit of zero quark masses, $m_q \to 0$, the dynamical problem in the semiclassical approximation to QCD reduces to a single variable and the LF Hamiltonian equation becomes the wave equation~\cite{deTeramond:2008ht, Brodsky:2014yha}
\begin{align} \lb{LFHWE}
\left(-\frac{d^2}{d\zeta^2} 
- \frac{1 - 4 L^2}{4 \zeta^2}+ U(\zeta) \right)  \phi(\zeta) = M^2 \phi(\zeta),
\end{align}
where $\zeta$ is the invariant physical transverse separation between partons in the light front, and $L$ is the orbital angular momentum in the transverse LF plane. Eq.~\req{LFHWE} is a frame-independent relativistic equation with a structure similar to that of Eq.~\req{AdSWE}. It leads to an exact mapping of AdS to the LF coordinates, provided that $z$ is identified with $\zeta$,  $z = \zeta$, and the AdS parameter $\nu$ with $L$~\cite{Brodsky:2006uqa, deTeramond:2008ht}.  The effective confinement potential $U(\zeta)$ in the Hamiltonian equation~\req{LFHWE} acts on the valence state and comprises all interactions, including those of the higher Fock states. It is obtained from~\req{Ugz} by mapping the gravitational coupling to LF physics at the Minkowski boundary of AdS space, $g_{s}(z) \to g_{s}(\zeta)$, thus
\begin{align} \lb{Ugzeta}
U(\zeta) =  g_{s}(\zeta) \frac{1}{\zeta^{2J -3}}  \frac{d}{d \zeta} \left(\zeta^{2J - 3} \frac{d}{d \zeta} g_{s}^{-1}(\zeta) \right),
\end{align} 
for $d = 4$, with $J$ the total hadron spin.

The physical strong coupling $\alpha_s$, measured at the momentum transfer $Q^2 = - q^2 > 0$, is the Fourier transform of the  LF coupling $g_s(\zeta)$ written in terms of the invariant separation~$\zeta$ 
\begin{align} \lb{alphavphi}
\alpha_{s}(\zeta) = \frac{g^2_{s}(\zeta) }{4 \pi} =  \frac{1}{4 \pi} e^{- \varphi(\zeta)},
\end{align}
integrated in the transverse LF plane. It is thus given by the Bessel transform~\cite{Brodsky:2010ur}
\begin{align} \lb{alphaint}
\alpha_s(Q^2) \propto \int_0^\infty  \! \zeta d\zeta J_0(\zeta Q) \,  \alpha_{s}(\zeta),
\end{align}
where a proportionality constant,  irrelevant to the form of the potential $U$, is left unspecified.

\subsection{Superconformal symmetry and the AdS coupling}   

An important additional step follows from the mechanism introduced by de Alfaro, Fubini, and Furlan~\cite{deAlfaro:1976vlx} to break the conformal symmetry in the Hamiltonian while keeping the action conformal invariant.   When extended to LF holographic QCD~\cite{Brodsky:2013ar}, this procedure leads to the appearance of a mass scale in the LF Hamiltonian. When extended to superconformal symmetry~\cite{Fubini:1984hf, Akulov:1983hjq} this mechanism uniquely determines the confining interaction $U$~\cite{deTeramond:2014asa, Dosch:2015nwa} and consequently the dilaton profile that breaks the conformality in the AdS action~\req{SAdS}. It also leads to hadronic supersymmetry among mesons, baryons, and tetraquarks~\cite{Dosch:2015nwa, Brodsky:2016yod}.  

For mesons, the confinement potential is~\cite{Dosch:2015nwa}
\begin{align} \lb{Uzeta}
U(\zeta) = \kappa_0^4 \zeta^2 + 2 \kappa_0^2 (J - 1),
\end{align}
and the coupling $g_s(\zeta)$ follows from integrating~\req{Ugzeta} for the expression~\req{Uzeta}.  We find~\cite{Dosch:2016zdv, deTeramond:2023qbo}
\begin{align}
g_s(z) =  \frac{1}{A\, e^{\kappa_0^2 z^2/2} + B \, \Gamma\left(2-J, \kappa_0^2 z^2\right) e^{-\kappa_0^2 z^2/2}},
\end{align}
where $\Gamma(a,b)$ is the incomplete Gamma function.  In the semiclassical approximation (ignoring the backreaction of the metric), the dilaton profile should depend only on the modification of AdS space, therefore independent of $J$. This condition implies that $B = 0$, and  since the coupling $g_s(\zeta)$ is defined up to a constant, we set $A = 1$. Thus the solution 
\begin{align} \lb{vphz}
    g_s(z) = e^{ - \kappa_0^2 z^2/2},
\end{align}
determined by the superconformal algebraic symmetry, where the value of the mass scale $\kappa _0 \simeq 0.5$ GeV is fixed by hadron spectroscopy~\cite{Brodsky:2014yha}. From~\req{alphavphi} $\alpha_{s}(\zeta) \sim e^{- \kappa^2 \zeta^2}$ and therefore
\begin{align} \lb{alphaholog}
\alpha_s(Q^2) = \alpha_s(0) e^{- Q^2 / 4 \kappa_0^2},
\end{align}
the result from Ref.~\cite{Brodsky:2010ur}. It leads to an IR fixed point $\alpha_s(0)$ whose value depends on the scheme chosen to extract $\alpha_s(Q^2)$. It is clear from~\req{Ugzeta} that the confinement potential is scheme independent, as it should.

\subsection{The Bjorken $g_1$ spin sum-rule scheme \lb{g1}}

Here, as in~\cite{Brodsky:2010ur, deTeramond:2024ikl}, we define the QCD running coupling as the effective charge~\cite{Grunberg:1980ja, Grunberg:1982fw} 
from the Bjorken spin-sum rule~\cite{Bjorken:1966jh, Bjorken:1969mm}  denoted $\alpha_{g_1}$
\begin{align} \lb{Bjsr}
\frac{\alpha_{g_1}(Q^2)}{\pi} = 1 - \frac{6}{g_A} \int_0^{1 - \epsilon} dx \left[g_1^p(x, Q^2) - g_1^n(x,Q^2)\right],
\end{align}
which is a physical observable.  It leads to a well-defined value for the IR fixed point $\alpha_{g_1}(0) = \pi$  and is well measured in the IR and UV domains~\cite{SpinMuon:1993gcv, SpinMuonSMC:1994met, SpinMuonSMC:1997voo, SpinMuon:1995svc, SpinMuonSMC:1997mkb, COMPASS:2010wkz, E143:1994vcg, E143:1995rkd, E142:1996thl, E143:1995clm, E143:1996vck, E154:1997xfa, E154:1997ysl, E143:1998hbs, E155:1999pwm, E155:2000qdr, HERMES:1998pau, HERMES:2000apm, HERMES:2002gmr, Deur:2004ti, Deur:2008ej, Deur:2014vea, Deur:2021klh}.

Transforming from the $g_1$ scheme to any RS, e.g. $\overline{\rm MS}$ or MOM, can be done in the UV domain using the Bjorken sum rule~\cite{Deur:2016tte} or commensurate scale relations~\cite{Brodsky:1994eh, DiGiustino:2024zss}. In particular, the effective charge $\alpha_{g_1}$ can be obtained from the $\overline{\rm MS}$ results,
\begin{align} \lb{MSg1}
\frac{\alpha_{g_1}(Q^2)}{\pi}  =   \frac{\alpha_{\overline{\rm MS}} (Q^2) }{\pi}+   a_1 \frac{\alpha^2_{\overline{\rm MS}}(Q^2)}{\pi^2}  +  a_2  \frac{\alpha^3_{\overline{\rm MS}}(Q^2)}{\pi^3}  +   \cdots,
\end{align} 
where the coefficients $a_i$ are known to fourth order in perturbation theory~\cite{Kataev:1994gd, Kataev:2005hv, Baikov:2010je}.

Many choices for the definition of $\alpha_s$ are in fact available with consistent UV behavior, but significantly distinct in the IR~\cite{Deur:2016tte}. In contrast to the universally accepted quantum electrodynamics (QED) coupling definition as an effective charge~\cite{Gell-Mann:1954yli}, a consensus has yet to be reached regarding which definition should be adopted for $\alpha_s.$ Ultimately, such definition must be decided based on its usefulness~\cite{Deur:2023dzc}. The new expression for the potential in terms of $g_s(z)$, Eq.~\req{Ugz}, demonstrates the pertinence of adopting the effective charge definition, Eq.~\req{Bjsr}, for the canonical definition of the QCD coupling, consistent with QED. This definition provides a well-behaved coupling in the IR, whose interpretation is clear and follows the standard understanding of a coupling, and is computable with nonperturbative approaches. Yet, its usefulness has been limited in practice limited to a few predictions because its connection to the QCD potential had not been explicitly established.  Although formally derived in the IR domain, Eq.~\req{Ugz} provides compelling motivation for promoting the effective charge as the canonical definition for the QCD coupling. In other words, with Eq.~\req{Ugz}, knowledge of the coupling behavior,  by means of experimental data~\cite{Deur:2022msf} or explicit calculations~\cite{Brodsky:2010ur, Cui:2019dwv}, provides the potential $U(z)$ from which, in turn, many QCD observables can be computed.

\subsection{IR-UV non-analytic point-matching procedure \lb{Q0}}

The work in~\cite{Brodsky:2010ur} was subsequently extended by matching, at a single point $Q_0$, the holographic Gaussian expression~\req{alphaholog} with the perturbative QCD running coupling $\alpha_s(Q^2)$ and its $\beta$-function, $\beta(\alpha_s)$.  This procedure gives a good description of the IR domain and, by construction, of the UV data at all scales, since the actual matching can be performed at higher loops in the perturbative domain~\cite{Deur:2014qfa, Deur:2016cxb}. This simple non-analytic procedure allowed us to determine the matching point $Q_0 = 1.06 \pm 0.15$ GeV~\cite{Deur:2016opc}, which was used as the initial scale of perturbative evolution of QCD in subsequent HLFQCD predictions of parton distributions \cite{deTeramond:2018ecg, Liu:2019vsn, deTeramond:2021lxc}, and the scale-dependent Pomeron analysis in~\cite{Dosch:2022mop}. However, the point-matching procedure does not provide further insight into the transition domain or confinement mechanisms. Recently, the holographic QCD point-matching procedure~\cite{Brodsky:2010ur, Deur:2014qfa, Deur:2016cxb} has been used in the study of the holographic running coupling from the Ricci flow~\cite{Cancio:2024apu}. The matching procedure has also been updated by enforcing continuity up to the second derivative using the holomorphic coupling approach~\cite{Ayala:2024ghk}.

\section{IR-UV transition domain \lb{IRUV}}

Henceforth, we adopt the $g_1$ scheme, since the Bjorken sum is, to our knowledge, the best-suited observable to define $\alpha_{\rm eff}$ in the IR~\cite{Deur:2009zy}.  Furthermore, since the strong coupling in the $g_1$ scheme is a physical observable, it allows us to use analyticity properties as a guiding principle to describe the transition regime from the IR to the UV domains. In fact, in~\cite{deTeramond:2024ikl} an analytic expression for the effective coupling constant was proposed which describes essential features of the short and long distance behavior of QCD:
\begin{align} \lb{alphaeff}
\alpha_{\rm eff}(Q^2) = \alpha_{\rm eff}(0)\,  \exp\left[-\int_0^{Q^2} \frac{du}{4 \kappa_0^2 +
u\, \ln\left(\frac{u}{\Lambda^2}\right)}\right] .
\end{align}
It incorporates  the IR fixed point $\alpha_{\rm eff}(0)$  and the nonperturbative confinement scale $\kappa_0^2$ defined in the holographic model. The scale $\Lambda$ is introduced by the need for renormalization in the perturbative domain.  The small and large $Q^2$ behavior of~\req{alphaeff} is
\begin{align} \lb{alphalim}
\alpha_{\rm eff}(Q^2) =
\Bigg\{ \begin{array}{rcl}  
& e^{- Q^2 /  4 \kappa^2},  &Q^2 \le 4 \kappa^2 , \\
& K / \ln\left(Q^2 / \Lambda^2 \right),     &Q^2 \gg 4 \kappa^2.
\end{array}  
\end{align}
where $K$ is a dimensionless constant computed in Appendix~\ref{asy}. We thus recover the holographic exponential dependence in the IR~\cite{Brodsky:2010ur} and the logarithmic behavior characteristic of asymptotic freedom in the UV~\cite{Gross:1973id, Politzer:1973fx}.

The effective coupling~\req{alphaeff} leads to a continuous transition from IR to UV, where the QCD iteration methods are applicable.
It is defined for $\kappa_0^2 > \Lambda^2 / 4 e$ and leads to the maximal analyticity relation
\begin{align} \lb{mar}
\kappa_0^2 = \frac{\pi}{8} \Lambda^2,
\end{align}
which follows from the study of the flow of singularities of the effective coupling~\req{alphaeff} in the $Q^2$-complex plane~\cite{deTeramond:2024ikl}.
The strong coupling~\req{alphaeff} provides a new analytic framework to extend the holographic IR results to the near-perturbative transition domain, where it provides an accurate description of the available data in terms of a unique scale.

\section{Large $Q^2$ behavior \lb{LQ2}}

\subsection{Large-$Q^2$ behavior of the effective coupling}

The large $Q^2$ behavior of the effective coupling~\req{alphaeff} is derived in Appendix~\ref{asy}. It is given by
\begin{align} \lb{aleffLQ2}
\alpha_{\rm eff}(Q^2) &= \alpha_{\rm eff}(0)\, \frac{K_w(\rho)}{\ln\left(\rho\,\frac{Q^2}{4 \kappa_0^2}\right)}  
\left(1 + \mathcal{O} \left(\frac{4 \kappa_0^2}{\rho Q^2} \right)\right),
\end{align}
where
\begin{align} \lb{rhodef}
\rho = \frac{4 \kappa_0^2}{\Lambda^2},  \quad  \quad  \frac{1}{e} < \rho \le \frac{\pi}{2}.
\end{align} 
The variable $w$, $0 < w < Q^2 / 4 \kappa_0^2$, is an intermediate integration point in the integral in~\req{alphaeff}, and $K_w(\rho)$ is computed numerically. It is given by the expression
\begin{align} \lb{Krho}
K_w(\rho) = \ln(\rho w) \exp{ \left( - \int_0^w \frac{dv}{ 1+v\, \ln(\rho\,v) } \right)} \left(1 + \mathcal{O} \left(\frac{1}{w}\right)  \right),
\end{align}
which is well defined for $\rho > 1/e  \simeq 0.3679$. As we show in Appendix~\ref{asy}, $K_w(\rho)$ has a fast convergence to its asymptotic value for maximal analyticity~\req{mar}, $\rho = \pi/2$,
\begin{align} \lb{Krhohalf}
K_{w \to \infty}(\pi/2)= 0.285785,
\end{align}
thus, it mainly depends on the parameter $\rho$~\req{rhodef} for $w$ sufficiently large.

\subsection{Large-$Q^2$ behavior of perturbation theory \lb{betaexp}}

The expression of the $\beta$-function in higher order perturbation theory for an $SU(N_c)$ gauge theory is given by the RGE~\cite{Gross:1973id, Politzer:1973fx} 
\begin{align} \lb{rge}
\beta(\alpha_s) &= \mu^2 \frac{\partial \alpha_s(\mu^2)}{\partial \mu^2}  \\
  &= - \beta_0 \alpha_s^2 - \beta_1 \alpha_s^3  -  \beta_2 \alpha_s^4  + \cdots,
\end{align}
where the coefficients $\beta_0, \beta_1, \beta_2, \cdots$ depend on the number of colors $N_c$ and  $n_f$, the number of active quark flavors at the renormalization scale $\mu$. The one and two-loop coefficients, $\beta_0$ and $\beta_1$, are independent of the renormalization scheme and are given by~\cite{ParticleDataGroup:2024cfk}
\begin{align}
\beta_0 &= \frac{11 N_c -2 n_f}{12 \pi},  \nn  \\
\beta_1 &=\frac{1}{24 \pi^2}\left[17 N_c^2-n_f \left(5 N_c + 3\frac{N_c^2-1}{2 N_c}\right)\right].  \lb{rgcf} 
\end{align}
From the recursive integration of~\req{rge} one obtains at the next to leading order (NLO) the scheme independent expression,
\begin{align} \lb{alphasNLO}
\alpha_s(Q^2) = \frac{1}{\beta_0   \ln \left(Q^2 / \Lambda_s^2 \right)}
 \left[1  - \frac{\beta_1}{\beta_0^2}\frac{\ln \ln \left(Q^2 /\Lambda_s^2 \right)}{ \ln \left(Q^2 / \Lambda_s^2 \right)} + \cdots \right] , 
 \end{align}
where the omitted higher-order scheme-dependent terms decrease as $1/ \ln^2 \left(Q^2/\Lambda_s\right)$~\cite{Deur:2016tte}. We have also identified the renormalization scale $\mu$ with the momentum transfer $Q$ in~\req{alphasNLO}. Close to the subtraction point  $\mu^2 \simeq Q^2$,  $Q^2$ becomes the virtuality for a given process.  Notice that $\Lambda_s$ refers to the QCD scale in a perturbative iteration, while $\Lambda$, without a subscript, refers to the effective scale that controls the logarithmic dependence~\req{alphaeff}.  It was shown in~\cite{deTeramond:2024ikl} that, to a good approximation, $\Lambda_s = \Lambda$ for maximal analyticity.

\section{Scale dependent hadronic strength \lb{fu}}

The boundary condition at infinity modifies the scaling properties of the effective coupling $\alpha_{\rm eff}$ in the ultraviolet domain, and leads to a $Q^2$-flow for the confinement strength, $\kappa$, constrained by the asymptotic symmetry and the analytic properties described in~\cite{deTeramond:2024ikl}.

\subsection{Scaling properties of the effective coupling}

To properly incorporate the scale dependence of $\kappa$  into the effective coupling $\alpha_{\rm eff}$,  it is necessary to account for its evolution with  $Q^2$ by introducing a dimensionless holomorphic function $f(Q^2)$ in the complex plane, except at the unitarity branch cuts. A consistent procedure amounts to rescaling the variable $u$ in the integral~\req{alphaeff} according to
\begin{align} \lb{ufu}
u \to \frac{u}{f(u)},
\end{align}
which is equivalent to perform the rescaling $\kappa_0^2 \to \kappa_0^2 \, f(Q^2)$, thereby embedding the scale dependence directly into the effective coupling. 

From~\req{alphaeff} and~\req{ufu} we obtain
\begin{align} \lb{aluf}
\alpha_{\rm eff}\left(Q^2, f(Q^2) \right) = \alpha_{\rm eff}(0)\,  \exp\left(-\int_0^{Q^2} \frac{d(u / f(u))}{4 \kappa_0^2 + \left(u/f(u)\right) \, \ln\left(u / f(u) \Lambda^2\right)}\right),
\end{align}
with $\alpha_{\rm eff} \left(0,f(0)\right) \equiv \alpha_{\rm eff}(0)$, the IR fixed point value.  We can also express  Eq.~\req{aluf} in terms of a scale dependent $\kappa$  defined by
\begin{align} \lb{kfq}
\kappa^2(Q^2) = \kappa_0^2 \, f(Q^2), \quad {\rm where} ~~ f(Q^2) \to 1, ~~ {\rm for} ~~ Q^2 \to 0, 
\end{align}
and its UV behavior as $Q^2 \to \infty$ is determined by the structure of the gauge group at the asymptotic boundary. 
From~\req{rhodef} and~\req{aluf} we find
\begin{align} \lb{aluk}
\alpha_{\rm eff}\left(Q^2, \kappa (Q^2) \right)  = \alpha_{\rm eff}(0) \,  \exp\left(-\int_0^{Q^2}  \frac{\kappa^2(u) \partial_u (u / \kappa^2(u)) \, du }{4 \kappa^2(u) + u \, \ln\left(\rho\, u /4 \kappa^2(u) \right)} \right) .
\end{align}
The functional form of $\kappa(Q^2)$ must be chosen to smoothly interpolate between the IR and UV regimes,  such that  the effective coupling $\alpha_{\rm eff}(Q^2)$ preserves the analytic structure found in~\cite{deTeramond:2024ikl}. This requires that both $\kappa(Q^2)$ and $\alpha_{\rm eff}(Q^2)$ remain holomorphic in the complex $Q^2$ plane, except at the physical cuts associated with the heavy-quark thresholds for $Q^2 < 0$, and the flow of singularities studied in~\cite{deTeramond:2024ikl} for $Q^2 > 0$.

By further changing the integration variable	
\begin{align}
v = \frac{u}{4 \kappa^2(u)},
\end{align}
we can be express Eq.~\req{aluk} as
\begin{align} \lb{alrhov}
\alpha_{\rm eff}\left(Q^2, \kappa (Q^2) \right) = \alpha_{\rm eff}(0) \,  \exp\left(- \int_0^{Q^2/4 \kappa^2(Q^2)} \!  \! \! \frac{dv}{1+v\, \ln(\rho \, v) }\right).
\end{align}
Thus, the introduction of a scale-dependent $\kappa$ as given by~\req{kfq}, maintains, by construction, the flow of singularities described in~\cite{deTeramond:2024ikl} and, therefore, the maximal analyticity condition~\req{mar}, since the value of $\rho$~\req{rhodef} is invariant under~\req{ufu}.

The large $Q^2$ behavior of the effective coupling follows from the results in Appendix~\ref{asy} by modifying the upper limit of integration from~\req{alrhov} without further changes. We thus obtain at large $Q^2$
\begin{align} \lb{aleffLQ}
\alpha_{\rm eff}\left(Q^2, \kappa(Q^2) \right) = \alpha_{\rm eff}(0)\, \frac{K_w(\rho)}{\ln\left(\rho\,\frac{Q^2}{4 \kappa^2(Q^2)}\right)} 
\left(1 + \mathcal{O} \left(\frac{4 \kappa^2(Q^2)}{Q^2} \right)\right),
\end{align}
where $K_\omega(\rho)$ is given by~\req{Krho}.

\subsection{Scaling properties of the effective $\beta$-function}

From the scale dependent expression of the effective coupling~\req{aluf} we compute its $\beta$-function,
\begin{align} \lb{betafu} 
\beta(\alpha_{\rm eff}) &=  Q^2 \,  \frac{\alpha_{\rm eff}\left(Q^2, \kappa (Q^2) \right)}{d Q^2} \nn   \\
    &=  - Q^2 \frac{\kappa^2(Q^2) \, \partial_{Q^2} \left(Q^2 / \kappa^2(Q^2) \right)}{4 \kappa^2(Q^2) + Q^2 \ln\left(\rho\, Q^2 /4 \kappa^2(Q^2) \right)} \, \alpha_{\rm eff}\left(Q^2, \kappa (Q^2) \right),
\end{align}  
It reduces for $\kappa^2(Q^2) = \kappa_0^2$ to the expression of the $\beta$-function found in \cite{deTeramond:2024ikl}. At large  $Q^2$ we obtain from~\req{aleffLQ} and \req{betafu}
\begin{align} \lb{betaeffLQ}
\beta(\alpha_{\rm eff}) = - \beta^{\rm eff}_0(Q^2)   \alpha^2_{\rm eff}\left(Q^2, \kappa (Q^2) \right)
\end{align}
where
\begin{align} \lb{fdf}
 \beta^{\rm eff}_0(Q^2) = \frac{1}{\alpha_{\rm eff}(0) K_w(\rho)}  f(Q^2) \frac{d}{d Q^2} \left(Q^2 / f(Q^2)\right).
\end{align}
Eq.~\req{betaeffLQ} has a structure similar to the leading RGE result~\req{rge}, but with a $Q^2$-dependent factor $\beta_0^{\rm eff}(Q^2)$.

\section{Asymptotic gauge symmetry, number of flavors and UV scaling behavior \lb{UV}}

The actual comparison of the effective~\req{aleffLQ} and perturbative QCD couplings~\req{alphasNLO} requires a power-law scaling dependence of the dimensionless function $f(Q^2)$, to reach a finite value for their ratio in the asymptotic limit. Thus, for  $Q^2 \gg \kappa_0^2$  the scaling function $f(Q^2)$ should behave as
\begin{align} \lb{fUV}
f(Q^2) \to   (\sigma Q^2 )^r,  \  \quad r \ge 0,
\end{align}
where $r$ is the scaling exponent of $f$, and $\sigma > 0$ is a dimensionfull parameter, except for the trivial case $r=0$ . The power-law scaling leads to a relationship between $n_f$ and the exponent $r$ in the deep UV, on the one hand, and the value of the IR fixed point, on the other.

\subsection{Asymptotic boundary conditions for the effective coupling \lb{bcal}}
  
We incorporate the asymptotic symmetry in the model by imposing the boundary condition 
\begin{align} \lb{asycond}
\lim_{Q^2 \to \infty} \frac{\alpha_{\rm eff}\left(Q^2, \kappa (Q^2) \right)}{\alpha_s(Q^2)} = 1 ,
\end{align} 
at the asymptotic limit ${Q^2 \to \infty}$ for asymptotically free gauge theories. It leads to the result
\begin{align} \lb{mr}
\alpha_{\rm eff}(0) \beta_0 \, K_{w \to \infty}(\rho) = 1-r ,
\end{align}
where the value of the exponent $r$ in~\req{fUV} can be determined based on certain assumptions. Specifically, for a given value of the parameter $\rho$,  the scaling exponent $r$ depends on the value of the IR fixed point $\alpha_{\rm eff}(0)$, the Casimir group invariant $C_A = N_c$ and the number of flavors $n_f$ in $\beta_0$, independently of any dimensionful parameter.  By imposing maximal analyticity, $\rho = \pi/2$, we find for $N_c = 3$  and $n_f= 6$  in color $SU(3)$ the numerical result $1 - r = 0.500124$ in the Bjorken $g_1$ scheme, where $\alpha_{\rm eff}(0) = \pi$. This agrees with the exponent $r = 1/2$, from heavy-quark bound state systems in holographic QCD~\cite{Gutsche:2012ez, Dosch:2016zdv, Nielsen:2018ytt}, the only possible solution of $r$ for $N_C=3$ and $N_f=6$ (see Eq.~\req{r} in Appendix~\ref{Ancnfr}.

\begin{figure}[h]
\begin{center}   
\includegraphics[width=8.6cm]{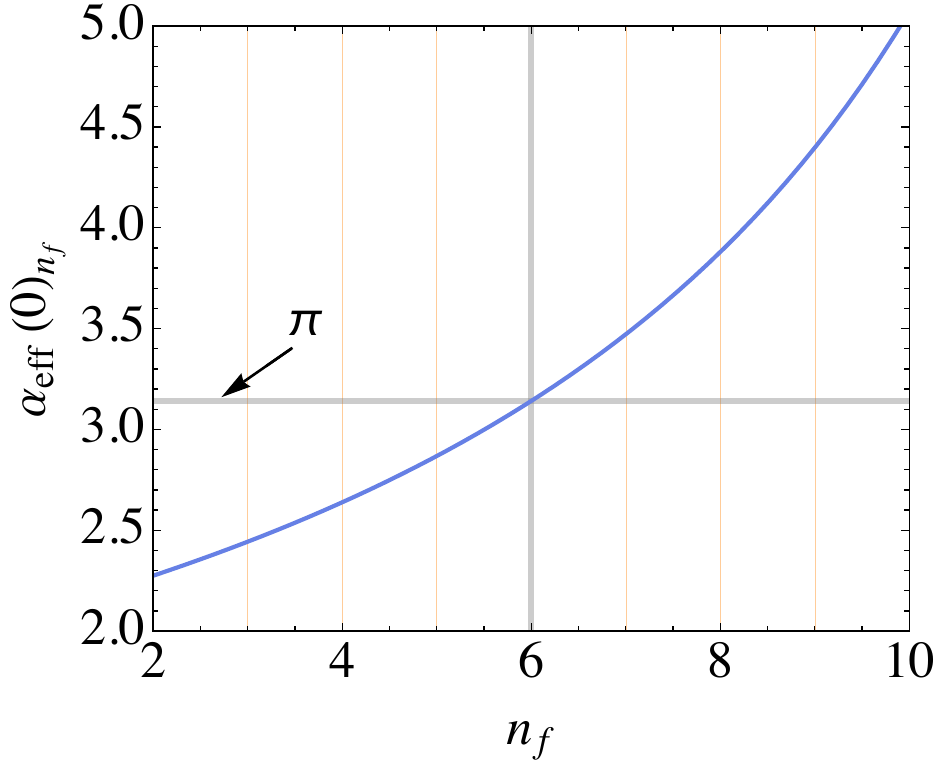} 
\caption{\lb{Fal0nf} The light blue curve represents the value of the infrared fixed point value $\alpha_{\rm eff}(0)_{n_f}$ from~\req{al0nfK} as a function of the number of flavors for $N_c = 3$ and  the maximal analyticity condition $\rho = \pi/2$. The value $\alpha_{\rm eff}(0) = \pi$ imposed by the Bjorken sum rule is depicted by the horizontal gray line: It crosses the vertical line precisely at $n_f = 6$.  This result is consistent with the power scaling $r= 1/2$ from heavy quark spectroscopy in holographic QCD.}
\end{center}
\end{figure}

Alternatively, we can start with the exponent $r =1/2$ following~\cite{Gutsche:2012ez, Dosch:2016zdv, Nielsen:2018ytt, Shuryak:1981fza, Isgur:1991wq}, leaving the number of flavors $n_f$ and the value of the IR fixed point unspecified. From~\req{mr}, 
\begin{align} \lb{al0nfK}
\alpha_{\rm eff}(0)_{n_f} = \frac{2 \pi}{(11 - \frac{2}{3} n_f) K_{w \to \infty}(\pi/2)},
\end{align}
for $N_c$ = 3 and maximal analyticity $\rho= \pi/2$.  We show in Fig.~\ref{Fal0nf} that the IR fixed point $\alpha_{\rm eff}(0)=\pi$, imposed by the Bjorken sum rule~\cite{Bjorken:1966jh,  Bjorken:1969mm}, requires six flavors in the deep UV, if we demand $r=1/2$. Thus, fixing $n_f = 6$ and $\alpha_{\rm eff}(0)=\pi$  we can evaluate $K_w(\rho)$~\req{Krho} with good precision. From~\req{al0nfK}
\begin{align} \lb{K27}
K_{w \to \infty}(\pi/2) = \frac{2}{7}  = 0.285714 \cdots,
\end{align}
compared with the value from numerical integration 0.285785.

The analytic structure of $\alpha_{\rm eff}(Q^2)$, determined by the scaling exponent $r$  in~\req{fUV}, is related to the number of colors and flavors. The power $r = 1/2$, which was previously observed in~\cite{Gutsche:2012ez, Dosch:2016zdv, Nielsen:2018ytt}, is very satisfactory for several reasons: It leads for $SU(3)$ color to the IR fixed point $\alpha_{\rm eff}(0)=\pi$, and coincides with $n_f = 6$ if we set $r=1/2$  (Fig.~\ref{Fal0nf}). 

From \req{rgcf},  \req{mr} and~\req{K27}  we find for $\alpha_{\rm eff} = \pi$ and maximal analyticity, $\rho = \pi/2$, the relation
\begin{align} \lb{ncnfr}
\frac{11 N_c - 2 n_f}{42} = 1 - r
\end{align}
in the asymptotic limit. For asymptotically free gauge theories $11 N_c - 2 n_f > 0$, thus the upper bound $r < 1$. Other values of $n_f$,  $N_c$ and $r$ are also compatible with the asymptotic boundary condition~\req{asycond}. This is discussed in Appendix~\ref{Ancnfr}.

\subsection{Asymptotic boundary conditions for the effective $\beta$-function \lb{bcbt}}

The embodiment of the gauge symmetry at the asymptotic boundary discussed above can be readily introduced in terms of the leading RGE results for the $\beta$-function. Indeed, comparing the effective $\beta$-function~(\ref{betaeffLQ}, \ref{fdf}) with the RGE expansion of the QCD $\beta$-function in powers of $\alpha_s$~\req{rge}, lays out the condition
\begin{align}
 f(u) \frac{d}{d u} \left(\frac{u}{f(u)}\right) = C,
\end{align}
with $C$ a constant. It has the solution
\begin{align} \lb{fur}
f(u) = \sigma u^r, \quad {\rm with} \quad r = 1 - C. 
\end{align}
Substituting the power solution~\req{fur} in \req{fdf} we obtain
\begin{align}
 \beta^{\rm eff}_0 = \frac{1 - r}{\alpha_{\rm eff}(0) K_{w \to \infty}(\rho)}  ,
\end{align}
which we compare with the leading coefficient of the QCD $\beta$-function $\beta_0$~\req{rgcf} with six open flavors
\begin{align} \lb{b0}
\beta_0 =  \frac{7}{4 \pi} .
\end{align}
Thus, we recover the solution $r = 1/2$ obtained above for maximal analyticity, $\rho = \pi/2$,  $K(\pi/2) = 2/7$ from~\req{K27}, and  $\alpha_{\rm eff}(0) = \pi$, consistent with the results in Sect.~\ref{bcal}.

\section{Threshold dynamics and the IR-UV analytic flow \lb{exp}}

We begin by considering a simple functional form for the scaling function $f(u)$ introduced in Sect.~\ref{fu} 
\begin{align} \lb{firuv}
f(Q^2) = \left(1 + \sigma Q^2\right)^r,  ~~  \sigma > 0,
\end{align}
which incorporates the  IR and UV limits~\req{kfq} and~\req{fUV} 
\begin{align} \lb{fbc}
f(Q^2) =
\Bigg\{ \begin{array}{rcl}  
& 1,  &Q^2 \to 0, \\
&(\sigma Q^2)^r ,   &Q^2 \to \infty,
\end{array}  
\end{align}
respectively, and interpolates smoothly between both.  For the possible values of the scaling exponent $r$  in the interval $0 \le r < 1$ (see Appendix~\ref{Ancnfr}), $f(Q^2)$ in~\req{firuv} has a branch cut in the timelike domain of the complex plane $q^2 = - Q^2 > 0$, from $q^2 = 1/\sigma$ to $q^2 \to \infty$.  Since there are no overlapping singularities arising from this additional cut, our previous results for the flow of singularities in the spacelike domain $Q^2 > 0$ are not modified~\cite{deTeramond:2024ikl}. Consequently,  maximal analyticity is maintained by~\req{firuv}, independently of the value of~$Q^2$.

\begin{figure}[h]  
\includegraphics[width=12.8cm]{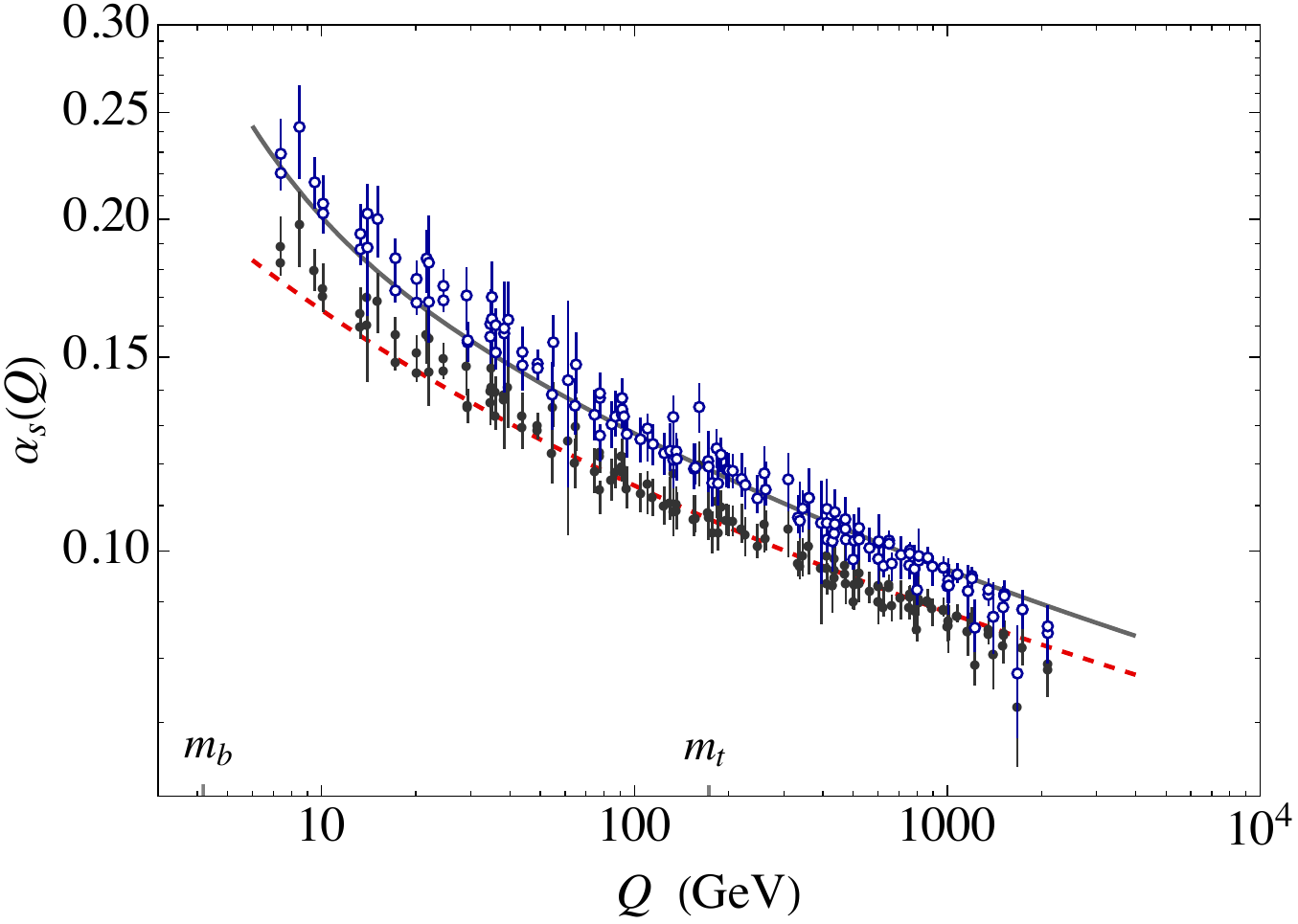}
\caption{\lb{alUV}  World data for $\alpha_{\overline{\rm MS}}$ as compiled by the PDG~\cite{ParticleDataGroup:2024cfk}  (filled black circles), is compared with the $\overline{\rm MS}$ data transformed to the $g_1$ scheme in the perturbative domain~\req{MSg1} (open blue circles).  The lower dashed red curve corresponds to the NLO approximation~\req{alphasNLO} for $n_f= 6$ and $\Lambda^{n_f =6}_{\overline{\rm MS}} =  87~{\rm GeV}$.  The results for the effective coupling~\req{aluf} (light gray curve) correspond to $\kappa_0 = 0.5345$ GeV~\cite{Sufian:2018cpj} and $\sigma = 0.015 ~ \rm{GeV}^{-2}$ in~\req{firuv} for $r=1/2$ and maximal analyticity. The bottom and top quark masses are also indicated for reference.}
\end{figure}

To extract the value of the parameter $\sigma$ in~\req{firuv}, we compare in Fig.~\ref{alUV} the  available data for $\alpha_{\overline{\rm MS}}$ in the perturbative domain~\cite{ParticleDataGroup:2024cfk},  with the same data set transformed into the $g_1$ scheme~\req{MSg1}.  The lower red dashed curve corresponds to the NLO scheme-independent approximation~\req{alphasNLO} for $n_f= 6$ and $\Lambda^{n_f =6}_{\overline{\rm MS}} = 87 ~ {\rm GeV}$. The UV data in Fig.~\ref{alUV}, specifically the data around 1000 GeV, are used to find the values of $\Lambda_{\overline{\rm MS}}$ and $\sigma$~\req{firuv}. The model results for the effective coupling~\req{aluf} (light gray curve) are also shown in this figure for the effective scaling function~\req{firuv}. We find the value $\sigma = 0.015~\rm{GeV}^{-2}$ for the value $\kappa_0 = 0.534$ GeV obtained from the hadron spectrum~\cite{Sufian:2018cpj}, $r = 1/2$ and the maximal analyticity condition~\req{mar}, namely $\rho= \pi/2$.

\subsection{Physical thresholds and UV sum rules}

The scaling function~\req{firuv} can be seen as an effective approximation that encodes the dynamics from the quark thresholds. Actually, one may expect a distinct contribution from each heavy quark flavor:
\begin{align} \lb{qth}
f(Q^2) &= \sum_{f=c,b,t} C_f \left(1 + \sigma_ {f}Q^2\right)^r \nn \\ 
   &= \sum_{f=c,b,t} C_f \left(1 + \frac{1}{m_f^2} Q^2\right)^r , \quad C_f \ge0,
\end{align}
with branch cuts in the timelike domain, $q^2 = - Q^2$, from $q^2 = m_f^2$ to $q^2 \to \infty$.  Only the heavy quarks $c$, $b$ and $t$ contribute to the sums in~\req{qth} since the ultra-relativistic $u$, $d$ and $s$ light quarks are a component of the confinement dynamics, thus embedded in the hadron scale $\kappa_0$. This distinction is not so clear for the charm quark with a mass $m_c \simeq 1.3$ GeV. It is located near the boundary of the IR-UV transition domain at $Q^2 \simeq 4 \kappa_0^2 \simeq 1~{\rm GeV}^2$~\cite{Deur:2016cxb, deTeramond:2024ikl}. Consequently, the contribution from the charm threshold may be non-zero, but relatively smaller than the combined $b$ and $t$ contributions, $C_c \ll C_b + C_t$.

The boundary conditions~\req{fbc} in the IR and UV lead to the sum rules
\begin{align} 
\sum_f C_f &= 1, \nn \\
\sum_f C_f \sigma_f^r &= \sigma^r . \lb{fsum}
\end{align}
where $\sigma_f = 1 / m_f^2$. The system of equations~\req{fsum} results in two independent solutions, which we choose as the $b$ and $t$ coefficients, $C_b$ and $C_t$, expressed in terms of the masses of the heavy quarks, plus two independent parameters $\sigma$ and $C_c$. We find
\begin{align}
C_b &= \frac{\sigma^r  -  C_c \sigma_c^r  - (1 - C_c) \sigma_t^r}{\sigma_b^r  - \sigma_t^r},  \nn \\
C_t  &= - \frac{\sigma^r  -  (1 - C_c) \sigma_b^r - C_c \sigma_c^r}{\sigma_b^r  - \sigma_t^r}.  \lb{Cbt}
\end{align}

\begin{figure}[h] 
\includegraphics[width=8.4cm]{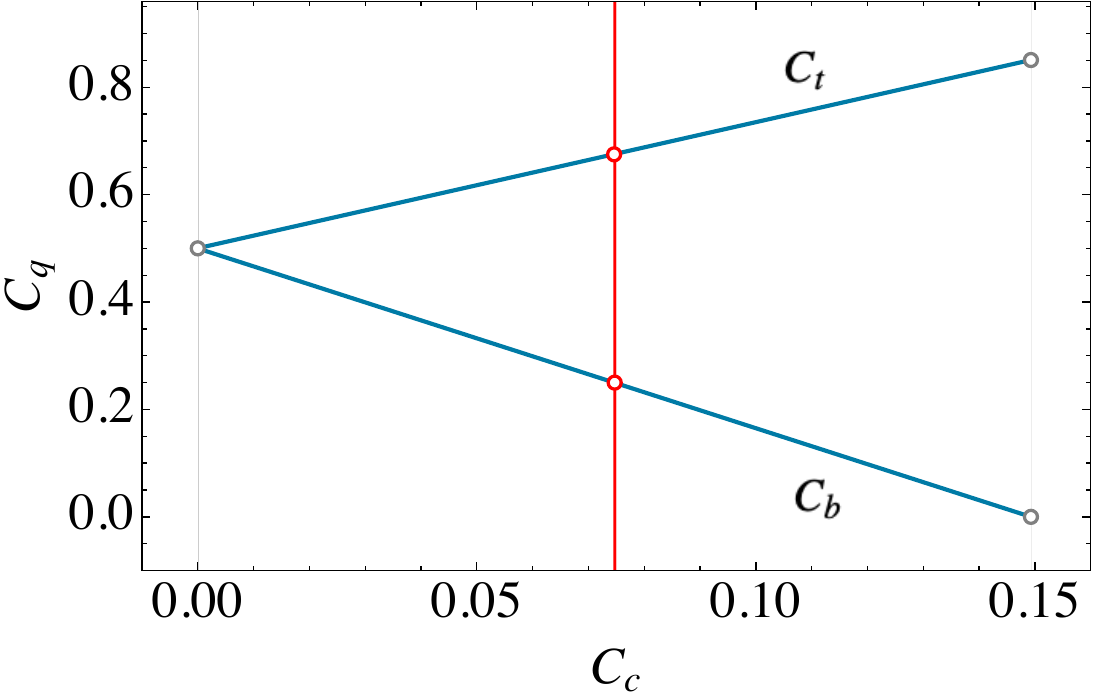}  \\ \vspace{20pt}
\includegraphics[width=8.8cm]{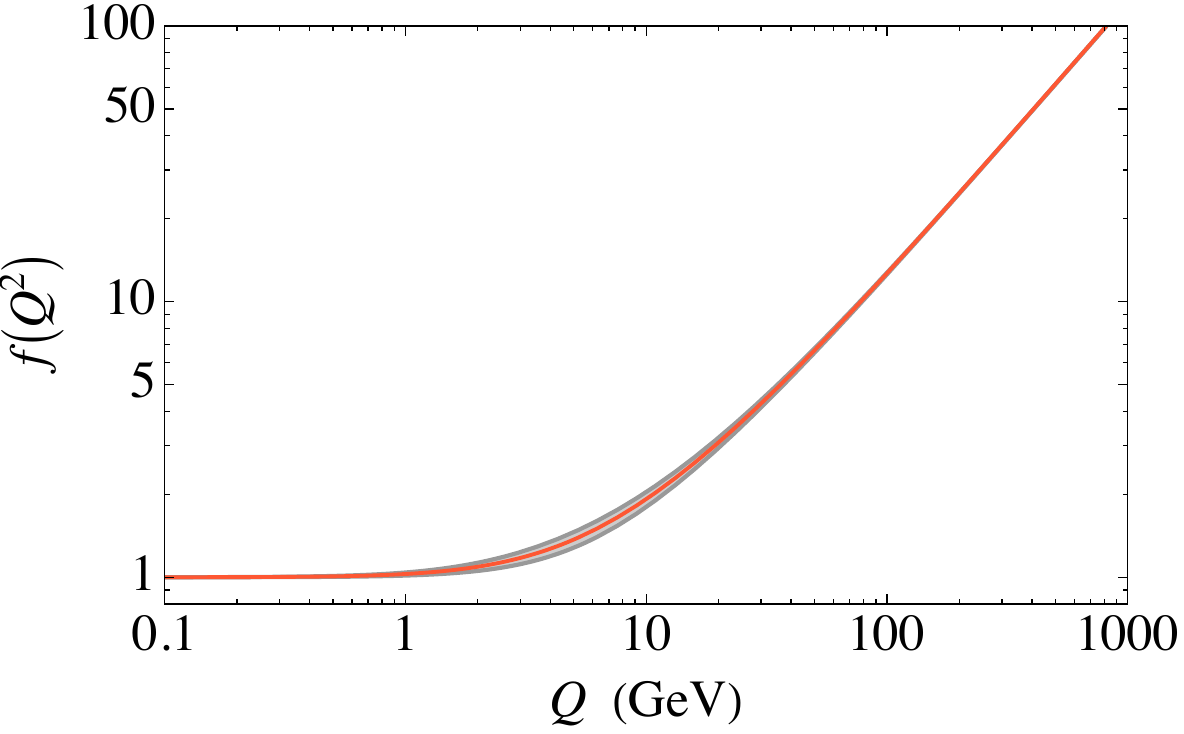}
\caption{\lb{locus} The locus of possible values of the charm, bottom and top weight coefficients, $C_c$, $C_b$ and $C_t$ (upper figure) is mapped into the locus of the possible trajectories of $f(Q^2)$ determined by the sum rules for $\sigma = 0.015~{\rm GeV}^{-2}$, as explained in the text.}
\end{figure}

If we fix the value of $\sigma $ from the fit to the UV data (Fig.~\ref{alUV}), $\sigma = 0.015~{\rm GeV}^{-2}$, we are left with the solution for $C_b$ and $C_t$ given in terms of the charm coefficient $C_c$. This is a parametric solution that determines the locus of the coefficients of $C_b$ and $C_t$ in terms of $C_c$,  as shown in Fig.~\ref{locus}. The upper curve corresponds to $C_t$,  $0.5  \le C_t \le 0.85$, and the lower curve corresponds to $C_b$, $0.5 \ge C_b \ge 0$.  Each value of  $C_c$ in the interval $0  \ge C_c \ge 0.15$ in the upper figure corresponds to one specific value for $C_b$ and $C_t$, which, in turn, are mapped in the lower figure to a specific trajectory of the scaling function $f(Q^2)$. For example, the value $C_c$ = 0.075 intercepts the curves for $C_b$ and $C_t$, pinpointing the values $C_b= 0.250$ and $C_t= 0.675$. This particular solution corresponds to the red curve in the lower figure. Likewise, since the red vertical line sweeps the interval from the initial to the final values of $C_c$, $0  \ge C_c \ge 0.15$, the locus of trajectories for $f(Q)$~\req{qth} is built in the lower figure, from the lowest gray curve corresponding to $C_s =0$, to the highest one corresponding to $C_s = 0.15$. The locus of the trajectories in Fig.~\ref{locus} corresponds to the values $\sigma_c=  0.62~{\rm GeV}^{-2}$,  $\sigma_b=  0.057~{\rm GeV}^{-2}$ and $\sigma_t= 3.4 \times 10^{-5} ~{\rm GeV}^{-2}$,  the inverse of the square masses of the charm, bottom and top quark, respectively $m_c =  1.27$ GeV, $m_b = 4.18$ GeV and $m_t= 172.76$ GeV.

\subsection{Locus reconstruction from the deep UV}

The locus of the scaling function $f(Q^2)$ in Fig.~\req{locus} indicates that all possible solutions converge to a unique trajectory in the deep UV, determined by the parameter $\sigma = 0.015~{\rm GeV}^{-2}$ extracted from the UV data. Furthermore, the equality of the coefficients $C_b$ and $C_t$ for $C_c = 0$ in Fig.~\ref{locus} indicates that $\sigma$ can be computed from the equality of the bottom and top coefficients, independently of the data fit. Thus, imposing the additional condition $C_b$ = $C_t$, for $C_c=0$, we can solve~\req{Cbt} exactly to determine the value of $\sigma$ from the point of common loci in Fig.~\ref{locus}. We find
\begin{align} \lb{sig}
\sigma =  2^{-1/r} \left(\sigma_b^r + \sigma_t^r \right)^{1/r}.
 \end{align}
 For $r = 1/2$ we can write~\req{sig} as
 \begin{align}
 \sigma = \frac{1}{4 \mu^2}, \quad  \frac{1}{\mu} = \frac{1}{m_b} + \frac{1}{m_t},
 \end{align}
with the value $\sigma = 0.01501~{\rm GeV}^{-2}$. Therefore, imposing the condition $C_b = C_t$, $C_c = 0$, one can determine the value of $\sigma$ relevant to UV behavior and, as a second step, reconstruct the locus, but independently of the data fitting procedure in Fig.~\ref{alUV}.

Having determined $\sigma$ precisely with the procedure described in this section, we compute the strong coupling at the $Z$-boson mass $M_Z = 91.2$ GeV. We find 
\begin{align}
\alpha_{\overline{\rm MS}}^{HLF}(M_Z) = 0.1161 \pm 0.0017.
\end{align}
Our prediction is compatible with the current value $\alpha_{\overline{\rm MS}}(M_Z) =  0.1180 \pm 0.0009$~\cite{ParticleDataGroup:2024cfk}, which includes lattice results.

\section{Comparison with the data at all scales \lb{allscal}}

\begin{figure}[h] 
\includegraphics[width=8.6cm]{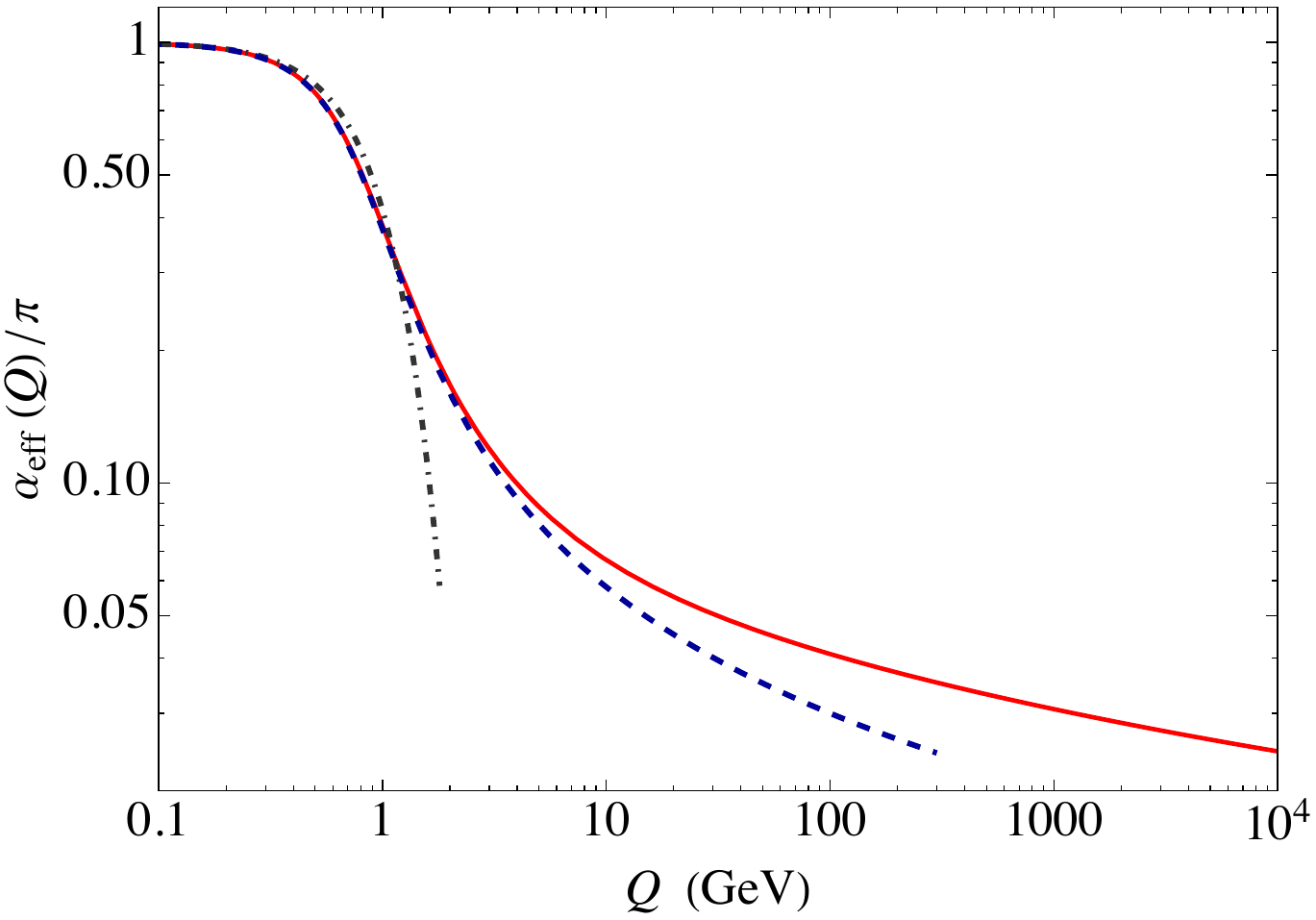}  
\caption{\lb{alcomp} The present model prediction from~\req{aluf} and~\req{qth} (red solid line) is compared with the IR-UV transition model results~\req{alphaeff} (blue dashed  curve) and the holographic exponential form~\req{alphaholog} (gray dot-dashed curve) for $\kappa_0 = 0.534$ GeV. The values for $C_q$ and $\sigma_q$ correspond to the red curve in Fig.~\ref{locus}, as explained in the text.}
\end{figure}

We compare the predictions from the present model~\req{aluf}  for the middle trajectory shown in Fig.~\ref{locus} (solid red line), with those from the IR-UV transition model~\req{alphaeff} (blue dashed curve) and the HLFQCD prediction~\req{alphaholog} (gray dot-dashed curve). We distinguish in Fig.~\ref{alcomp} three distinct domains: The IR nonperturbative holographic domain~\cite{Brodsky:2010ur} for $ 0 < Q^2 \leq 4 \kappa_0^2  \simeq 1~ {\rm GeV}^2$, the IR-UV transition domain studied in~\cite{deTeramond:2024ikl} for $4 \kappa_0^2 < Q^2  \lesssim 10^2 ~ {\rm GeV}^2$, and the UV extension described in this article for $10^2 ~ {\rm GeV}^2  \lesssim Q^2$, up to the deep UV domain. We notice that the value $4 \kappa_0^2 = 1.14 ~ {\rm GeV}^2$ is compatible with the value $Q_0^2 = 1.12  ~{\rm GeV}^2$~\cite{Deur:2016opc} from the IR-UV point-matching procedure described in Sect.~\ref{Q0}.

\begin{figure}[htbp]
\begin{center} 
\includegraphics[width=14.8cm]{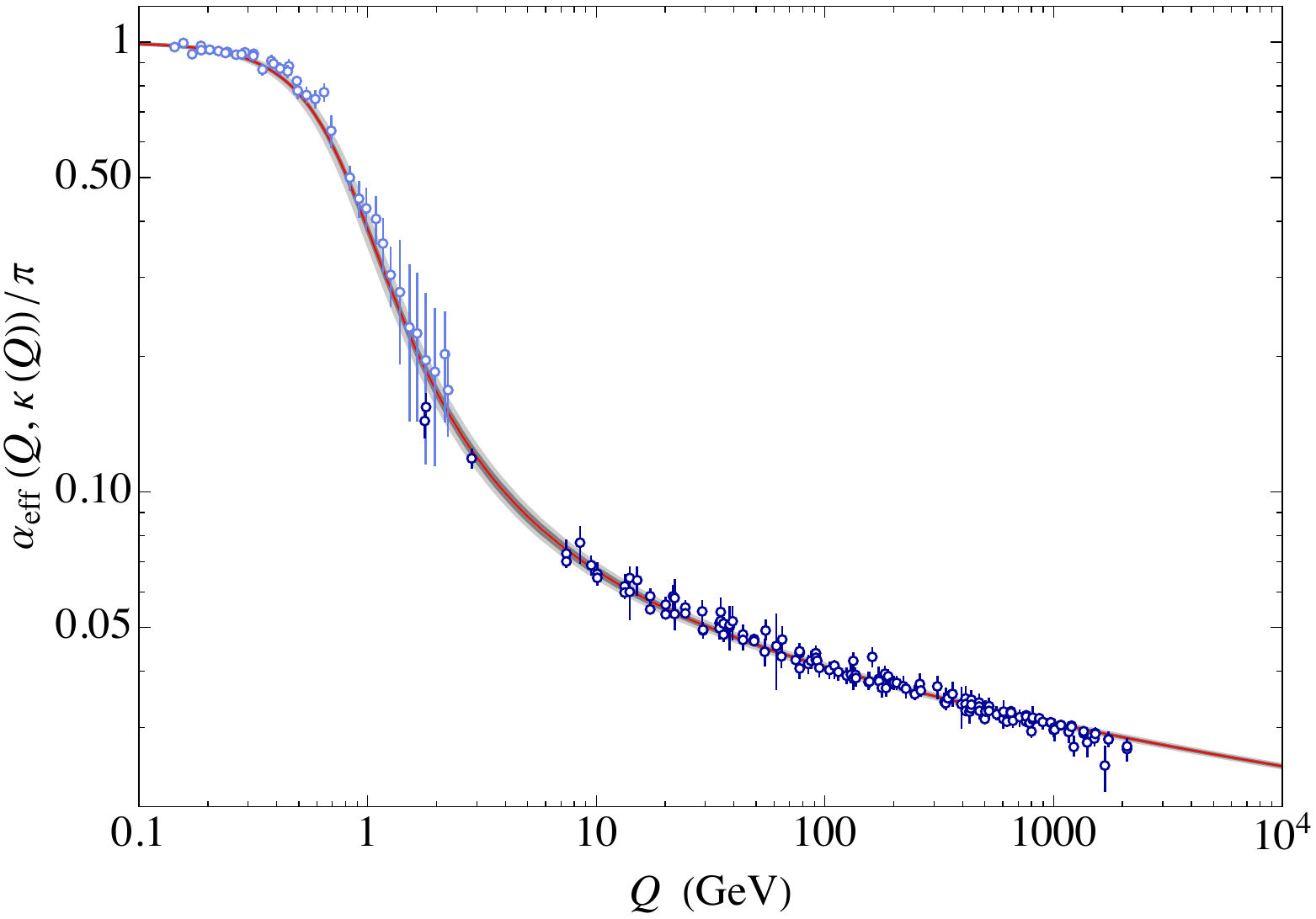}
\caption{\lb{alAll} The model predictions from~\req{aluf} and~\req{qth}  for $\kappa_0 = 0.534 $ GeV is compared with the the Bjorken spin sum rule data~\cite{SpinMuon:1993gcv, SpinMuonSMC:1994met, SpinMuonSMC:1997voo, SpinMuon:1995svc, SpinMuonSMC:1997mkb, COMPASS:2010wkz, E143:1994vcg, E143:1995rkd, E142:1996thl, E143:1995clm, E143:1996vck, E154:1997xfa, E154:1997ysl, E143:1998hbs, E155:1999pwm, E155:2000qdr, HERMES:1998pau, HERMES:2000apm, HERMES:2002gmr, Deur:2004ti, Deur:2008ej, Deur:2014vea, Deur:2021klh} (light-blue open circles), combined with the PDG data~\cite{ParticleDataGroup:2024cfk} transformed to the $g_1$ scheme (blue open circles). The red curve corresponds to the value $C_c$ = 0.075, $C_b= 0.250$ and $C_t= 0.675$. The lower and upper curves at the edge of the gray narrow band  for $C_c = 0$,  $C_b= 0.5$, $C_t = 0.5$ and $C_c= 0.15$, $C_b = 0$, and $C_t = 0.85$, respectively, define the boundaries of the possible solutions in Fig.~\ref{locus}. The values of $\sigma_c=  0.62~{\rm GeV}^{-2}$,  $\sigma_b=  0.057~{\rm GeV}^{-2}$ and $\sigma_t= 3.4 \times 10^{-5} ~{\rm GeV}^{-2}$ are computed from the heavy quark masses~\req{qth}. The light-gray wider band includes the uncertainty in the confinement scale $\kappa_0 = 0.534 \pm 0.025$ GeV obtained independently from the hadron spectrum.}
\end{center}
\end{figure}

We confront in Fig.~\ref{alAll} the model predictions with all available data, thus including the IR and transition domains.  Therefore, in addition to the PDG data~\cite{ParticleDataGroup:2024cfk} transformed to the $g_1$ scheme in the perturbative domain (blue open circles), we have included the data obtained directly from the Bjorken spin sum rule~\cite{SpinMuon:1993gcv, SpinMuonSMC:1994met, SpinMuonSMC:1997voo, SpinMuon:1995svc, SpinMuonSMC:1997mkb, COMPASS:2010wkz, E143:1994vcg, E143:1995rkd, E142:1996thl, E143:1995clm, E143:1996vck, E154:1997xfa, E154:1997ysl, E143:1998hbs, E155:1999pwm, E155:2000qdr, HERMES:1998pau, HERMES:2000apm, HERMES:2002gmr, Deur:2004ti, Deur:2008ej, Deur:2014vea, Deur:2021klh} (light blue open circles). The solid red curve  $\kappa_0 = 0.534$ GeV corresponds to the central red curve in Fig.~\ref{locus} for $f(Q)$, while the lower and upper curves at the edge of the gray narrow band correspond to the lower and upper curves for the locus of solutions for $f(Q)$ in the same figure. The wider light-gray band in Fig.~\ref{alAll} also includes the uncertainty in $\kappa_0$,  $\kappa_0 = 0.534 \pm 0.025$ GeV,  obtained independently from the light hadron spectrum~\cite{Sufian:2018cpj}. The prediction of the model for the effective coupling~\req{aluf}, with the heavy-quark scaling function~\req{qth}, gives a precise description of the data at all scales.

\begin{figure}[h] 
\includegraphics[width=8.6cm]{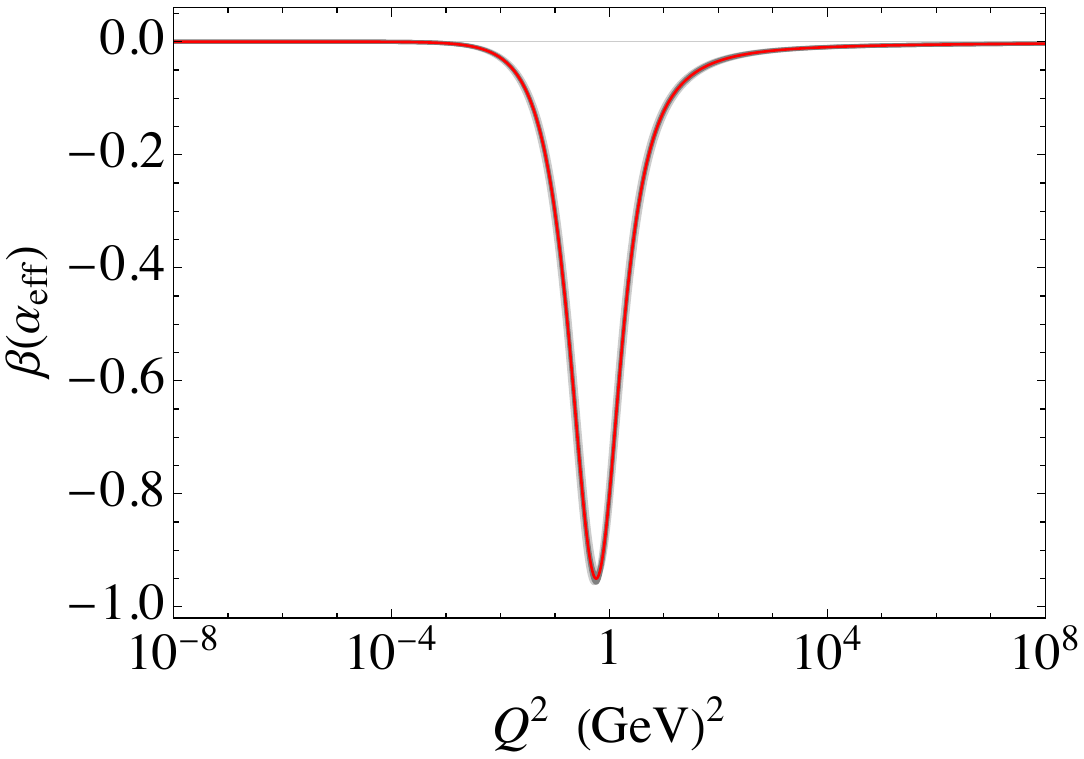}
\caption{\lb{betafig} The behavior of the $\beta$-function~\req{betafu} is computed using the same conventions as those in Fig.~\ref{alAll}. The figure illustrates the basic conformality of QCD, except at the confinement scale $Q^2 \simeq 4 \kappa_0^2 \simeq 1 \, {\rm GeV}^2$.}
\end{figure}

The $\beta$-function~\req{betafu} shown in Fig.~\ref{betafig} captures the essential conformal properties of QCD, except at the confinement scale $Q^2 \simeq 4 \kappa_0^2 \simeq 1 \, {\rm GeV}^2$. The  conventions used in this figure are the same as those in Fig.~\ref{alAll}.

\section{Conclusions and Outlook \lb{CaO}}

A unified nonperturbative analytical description of the QCD running coupling at all scales is an important step for understanding the confinement mechanisms in strongly coupled gauge theories and to bridge the gap between short-distance predictions and large-distance hadronic observables. To be useful, the running coupling must reflect in the UV the probability of gluon emission and, in the IR, the QCD potential from which hadronic phenomenology is derived. The latter connection is explicitly provided in the present article by Eq.~\req{Ugz}. In the nonperturbative analytic approach introduced in~\cite{deTeramond:2024ikl}, and extended here to the UV domain, there is a unique scale, the confinement scale $\kappa_0$, which also controls the decreasing logarithmic behavior in the UV.  This connection between the UV and IR domains becomes unique in the limit of maximal analyticity.

In QCD, in addition to the perturbative scale $\Lambda_s$, heavy-quark masses also influence the asymptotic behavior. Thus, it is not surprising that the massless theory described in~\cite{deTeramond:2024ikl} cannot reproduce the data in the UV region. To overcome this limitation, we have extended in this article the domain of applicability of~\cite{deTeramond:2024ikl} incorporating the symmetry group structure of asymptotically free gauge theories in the limit $Q^2 \to \infty$, with observable consequences in the UV domain.  In the context of holographic QCD, a symmetry at infinity corresponds to the symmetries of the dual quantum field theory at the AdS boundary at infinity, where QCD behaves as a conformal theory.

The specific scaling behavior in the deep UV, determined from asymptotic boundary conditions, requires a scale-dependent $\kappa$, the only dimensionful dynamical quantity in our model, with the same power scaling law in the UV as in previous heavy-quark spectroscopic results from holographic QCD~\cite{Gutsche:2012ez, Dosch:2016zdv, Nielsen:2018ytt}. This procedure requires introducing a specific scaling function, while maintaining the analytic structure of $\alpha_{\rm eff}$. For SU(3)$_C$, the deep UV boundary conditions require six flavors and an IR fixed point value $\alpha_{\rm eff}(0) = \pi$.

The scaling properties of the present model, are compatible with the introduction of branch cuts in the timelike domain of the complex plane, with no overlapping singularities in the spacelike domain~\cite{Horak:2023xfb}, therefore preserving the flow of singularities and the maximal analyticity properties in the spacelike domain~\cite{deTeramond:2024ikl}. Comparing with the QCD data for the running coupling in the UV, we are able to determine the locus of the threshold weighs for the charm, bottom, and top quarks, which converge to a unique solution in the deep UV, as shown in Fig.~\ref{locus}. One can also choose one specific solution,  for example equal bottom and top weights and zero weight for the charm, to determine the UV scaling behavior, without using the UV data. If we follow this alternative, we can accurately describe the strong coupling data at all scales, including the highest available data in the multi-TeV range~\cite{ParticleDataGroup:2024cfk},  in terms of one unified scale, the confinement scale $\kappa_0$, without any additional free parameter. In particular, we compare our prediction for the strong coupling with the current value of $\alpha_{\overline{\rm MS}}(M_Z)$~\cite{ParticleDataGroup:2024cfk}. We find its value in good agreement with our results.

The limit on the number of quark flavors, $n_f$, set by the nonperturbative approach discussed in this article is rather surprising and remarkable. The  $n_f=6$ limit follows from the incorporation of QCD asymptotic boundary conditions, combined with the well-determined value of the coupling’s IR fixed point.  It agrees with the bounds on the number of generations of quark and leptons, parametrized by the effective number of neutrino species, $N_{\rm eff}$, from the measurements of the primordial light elements in Big Bang Nucleosynthesis (BBN),  and the pattern of fluctuations in the Cosmic Microwave Background (CMB). The precise values are $N_{\rm eff} =2.88 \pm 0.27$ from combined BBN and CMB~\cite{Pitrou:2018cgg}, and  $N_{\rm eff} = 2.99 \pm 0.17$ from CMB combined with Baryon Acoustic Oscillations  in galaxy surveys~\cite{Planck:2018vyg}. Observations in collider experiments, also indicate that $n_f=6$. Those include the decay width of the $Z_0$ boson~\cite{ALEPH:2005ab} and the unitarity of the CKM matrix, where three generations are required~\cite{ParticleDataGroup:2024cfk}.

\appendix

\section{Large $Q^2$ behavior of the effective coupling \lb{asy}}

\begin{table}[htbp]
\caption{Numerical results for $K_w(\rho)$ show a rapid convergence for the maximal analyticity value $\rho = \pi/2$. For comparison, we present results for $\rho = \pi/4$, as an example. The second column indicates the corresponding $Q$ scale in GeV units for $4 \kappa_0^2  \simeq 1$. \lb{KrhoT}}
\begin{center}
\begin{tabular}{|c|c|c|c|}
\hline 
$w$  & $\sqrt{4\kappa_0^2 w}$  & $K_w(\pi/4)$ &  $K_w(\pi/2)$ \\
\hline \hline
   $10^2$  &10	&   0.0570879   &   0.285702\\
   $10^4$  & $10^2$	&   0.0571092   &   0.285785\\
   $10^6$  &$10^3$& 0.0571093   &   0.285785\\
\hline \hline
\end{tabular}
\end{center}
\end{table}

In this appendix we compute the large $Q^2$ behavior  of the effective coupling~\req{alphaeff}   
\begin{align} \lb{alphaeffA}
\alpha_{\rm eff}(Q^2) = \alpha_{\rm eff}(0)\,  \exp\left(-\int_0^{Q^2} \frac{du}{4 \kappa_0^2 + u\, \ln\left(\frac{u}{\Lambda^2}\right)}\right).
\end{align}
By changing the integration variable $v= u/4 \kappa_0^2$, the integral in~\req{alphaeffA} can be written as
\begin{align} \lb{F}
\int_0^{Q^2/4 \kappa_0^2} \! \! \frac{dv}{1+v\, \ln(\rho\,v) } 
&= \int_0^w\frac{dv}{1+v\, \ln(\rho\,v) } + \int_w^{Q^2/4 \kappa_0^2} \! \!  \frac{dv}{ 1+v\, \ln(\rho\,v) } \nn\\
&= \int_0^w\frac{dv}{1+v\, \ln(\rho\,v) } + \int_w^{Q^2/4 \kappa_0^2} \! \!  \frac{dv}{v\, \ln(\rho\,v)\left( 1+ \frac{1}{v\, \ln(\rho\,v)} \right)}\nn \\
&= \int_0^w\frac{dv}{1+v\, \ln(\rho\,v) } + 
\Big[\ln\left(\ln(\rho\,v)\right)\Big]^{Q^2/4 \kappa_0^2}_w -  \int_w^{Q^2/4 \kappa_0^2} \! \!  \frac{dv}{v^2 \ln^2(\rho v)} + \dots
\end{align}
where $ \rho = 4 \kappa_0^2/\Lambda^2$ and $w$ is an intermediate integration point. Following this, we obtain at  large $Q^2$
\begin{align} \lb{aleffLQA}
\alpha_{\rm eff}(Q^2) &= \alpha_{\rm eff}(0)\, \frac{K_w(\rho)}{\ln\left(\rho\,\frac{Q^2}{4 \kappa_0^2}\right)} 
\left(1 + \mathcal{O} \left(\frac{4 \kappa_0^2}{Q^2} \right)\right),
\end{align}
where 
\begin{align} \lb{KrhoA}
K_w(\rho) = \ln(\rho w) \exp{ \left( - \int_0^w \frac{dv}{ 1+v\, \ln(\rho\,v) } \right)} \left(1 + \mathcal{O} \left(\frac{1}{w}\right)  \right),
\end{align}
is well defined for $\rho > 1/e \simeq 0.3679$. As shown in Table~\ref{KrhoT}, $K_w(\rho)$ has a fast convergence, thus it basically depends on the value of $\rho$ for $w$ sufficiently large.

{\section{Higher-rank gauge groups, scaling, and asymptotic symmetry~\lb{Ancnfr}}

The relation between the asymptotic gauge group symmetry at the asymptotic limit, the number of open flavors $n_f$ and the scaling exponent $r$ in \req{fUV} follows from Eq.~\req{ncnfr}
\begin{align} \lb{ncnfrA}
\frac{11 N_c - 2 n_f}{42} = 1 - r,  \quad  0\le r < 1,
\end{align}
in the Bjorken sum rule scheme where $\alpha_{\rm eff}(0) = \pi$, with the value of $\rho$ in~\req{Krho} set by the analytic constraints of the model, $\rho = \pi/2$.

\begin{table}[htbp]
\caption{Number of colors and flavors, $N_c$ and $n_f$,  from the solutions of the  Diophantine equation~\req{ncnfrA}  for the possible power scaling values of  $r$ in the interval  $0 \le r < 1$. For $N_c = 3$ and $n_f=6$  there is only one solution $r = 1/2$. \lb{ncnfrT}}
\begin{center}
\begingroup
\setlength{\tabcolsep}{6pt} 
\renewcommand{\arraystretch}{1.1} 
\begin{tabular}{ |p{14pt}||p{14pt}|p{14pt}|p{14pt}|p{14pt}|p{14pt}|p{14pt}|p{14pt}|p{14pt}|p{14pt}|p{14pt}|p{14pt}|p{14pt}|p{14pt}|p{14pt}||}
 \hline 
 \multicolumn{1}{|c||}{$N_c$} &  \multicolumn{14}{|c|}{$n_f$} \\ 
 \hline \hline  
~  2     &     &	     &		&~4  &	 & 	&	&	& 	&~2	&~5	&~8	&	&   \\ 
~ \bf  3&     &\bf 6&	 &     &  &\bf 13&	&	&	&	&	&	&      &  \\ 
~ 4      & ~1& 	    &	~8	& 15	 &   &    & ~4	&~7	&10	&13	&16	&19	&       &  \\ 
~ 5      &	 & 17    &		&      &10	&24	&	&      &	&	&	&	&~8	&11\\ 
~ 6      & 12& 	    &	19	& 26  &     &	& 15 & 18	&21	&24	&27	&30	&	&    \\ 
~ 7      &      & 28  &		&       &21	& 35	&	&	&	&	&	&	&19	&22\\ 
$\cdots$ &    &        &	&	&	&	 &	&	&	&	&	&	&	&    \\ 
 \hline 
~ $ r $ & \, 0  & ~$\frac{1}{2} $& ~$\frac{1}{3} $ & ~$\frac{2}{3} $	&  ~$\frac{1}{6} $ &  ~$\frac{5}{6} $& ~$\frac{1}{7} $&~$\frac{2}{7} $&~$\frac{3}{7} $&~$\frac{4}{7} $&~$\frac{5}{7} $&~$\frac{6}{7} $&$\frac{1}{14} $&~$\frac{3}{14} $ \\
\hline\hline
\end{tabular}

\vspace{15pt}

\begin{tabular}{|p{14pt}||p{14pt}|p{14pt}|p{14pt}|p{14pt}|p{14pt}|p{14pt}|p{14pt}|p{14pt}|p{14pt}|p{14pt}|p{14pt}|p{14pt}|p{14pt}|p{14pt}||}
 \hline 
 \multicolumn{1}{|c||}{$N_c$} &  \multicolumn{14}{|c|}{$n_f$} \\ 
 \hline \hline  
~  2     &     &	        &	  &       &	  & 	&	&	&	& 	&~1	&~3	&~6	&~7	\\ 
~ \bf  3&~\bf3&~\bf 9&\bf12&\bf15&	  &	&	&	&	&	&	&	&	&	\\ 
~ 4      &     & 	       &	  &        &~2&~3&~5	& ~6	& ~9	&11	&12	&14	&17	&18 	\\ 
~ 5      &14 & 20      &23	  &  26  &	  &	 &	&	&      &	&	&	&	&	\\ 
~ 6      &     & 	       &	  &        &13& 14&16	& 17 & 20	&22	&23	&25	&28	&29	\\ 
~ 7      & 25& 31      &34	  &  37  &	  &	 &	&	&	&	&	&	&	&	\\ 
$\cdots$ &    &         &	  &	&	&	 &	&	&	&	&	&	&	&	\\ 
 \hline 
~ $ r $   & $\frac{5}{14} $& $\frac{9}{14} $ & $\frac{11}{14} $ & $\frac{13}{14} $ &  $\frac{1}{21} $& $\frac{2}{21} $&$\frac{4}{21} $&$\frac{5}{21} $&~$\frac{8}{21} $&$\frac{10}{21} $&$\frac{11}{21} $& $\frac{13}{21}$ &  $\frac{16}{21}$  &  $\frac{17}{21}$ \\
 \hline\hline
\end{tabular}

\vspace{15pt}

\begin{tabular}{ |p{14pt}||p{14pt}|p{14pt}|p{14pt}|p{14pt}|p{14pt}|p{14pt}|p{14pt}|p{14pt}|p{14pt}|p{14pt}|p{14pt}|p{14pt}|p{14pt}|p{14pt}||}
 \hline 
 \multicolumn{1}{|c||}{$N_c$} &  \multicolumn{14}{|c|}{$n_f$} \\ 
 \hline \hline  
~  2     &~9 &10  &	&      &	 & 	 &	&	&	& 	&	&	&	&	\\ 
~ \bf  3&     &     &	&      &~\bf1&~\bf2&~\bf4	&~\bf5&~\bf7&~\bf8&\bf10	&\bf11&\bf14&\bf16\\ 
~ 4      & 20&21 &	&	&       &      &	& 	& 	&	&	&	&	&	\\ 
~ 5      &     &     &7	&9    &12	 &13	 &15	&16	&18  &19	&21	&22	&25	&27	\\ 
~ 6      & 31& 32&	&	&       &      &	&      &  	&	&	&	&	&	\\ 
~ 7      &     & 	 &18	&20  &23	 &24	 &26	&27	&29	&30	&32	&33	&36	&38	\\ 
$\cdots$ &    &   &	&	&	 &	 &	&	&	&	&	&	&	&	\\ 
 \hline 
~$ r $ &$\frac{19}{21}$ & $\frac{20}{21} $ & $\frac{1}{42} $ & $\frac{5}{42} $ & $\frac{11}{42} $ & $\frac{13}{42} $& $\frac{17}{42} $& $\frac{19}{42} $& $\frac{23}{42} $& $\frac{25}{42} $& $\frac{29}{42} $& $\frac{31}{42} $&$\frac{37}{42} $& $\frac{41}{42} $ \\
 \hline\hline
\end{tabular}
\endgroup
\end{center}
\end{table}

Possible values of $r$ in the interval $0\le r < 1$ are found by restricting the solutions of~\req{ncnfrA} to integer values of $N_c$ and $n_f$. Specifically, this means considering the integer factorization of 42, which includes the factors 2, 3, and 7, as well as their products 6, 14, 21, and 42. This is equivalent to the solution of a linear diophantine equation, which is a linear equation in two or more unknowns with integer coefficients~\cite{Mordell:1969}. We find 42 possible values for $r$
\begin{equation} \lb{r}
r = 0, \frac{1}{2},  \frac{1}{3}, \frac{2}{3}, \cdots  , \frac{41}{42},
\end{equation}
shown in Table~\ref{ncnfrT} with the corresponding values for  $N_c$ and $n_f$.

The value of $r$ depends only on the difference $11 N_c - 2 n_f$ and not on $N_c$ and $n_f$ separately. For SU(3), for example, all combinations for 
$N_c=3+2 j, \; n_f= 6+ 11 j, \; j=0,1,2 \cdots$, lead to the same value $r = 1/2$  with three colors and 6 flavors, thus generating the corresponding column in Table~\ref{ncnfrT}
\begin{align} \lb{ncnfj}
(N_c, n_f) =  (3,6), (5,17), (7, 28), \cdots, (3+2 j,6+11 j).
\end{align}

It is interesting to note that, while the values of the leading coefficient of the $\beta$-function~\req{rgcf}  are the same for all the values in the table entry generated by~\req{ncnfj}, $\beta_0= 7 / 4 \pi$,  the higher coefficients can be drastically different from the ones of real QCD. For example,   $\beta_1= 13/ 8 \pi^2$  for $(N_c, n_f) = (3, 6)$, but the sequence
 \begin{align}
 \beta_1= + \frac{13}{8 \pi^2}, - \frac{51}{10 \pi^2}, - \frac{145}{8 \pi^2}, \cdots ,
 \end{align}
 increasingly diverges for the higher rank groups in the same column.  An interesting case occurs for constant scaling with $r = 0$. Here, $n_f=1$ corresponds to $N_c = 4$ and $\beta_0= 7/2 \pi$, twice the QCD value $\beta_0= 7/4 \pi$ for $n_f = 6$ and $r = 1/2$~\req{b0}.

\acknowledgments

The authors thank David d'Enterria and Klaus Rabbertz for providing the high-energy data compiled in the review~\cite{ParticleDataGroup:2024cfk}. This work is supported in part by the Department of Energy Contract No. DE-AC02-76SF00515. It is also supported in part by the US Department of Energy, Office of Science, and Office of Nuclear Physics under Contract No. DE-AC05-06OR23177. T. L. is supported by the National Natural Science Foundation of China under Grants No. 12175117 and No. 12321005 and Shandong Province Natural Science Foundation Grant No. ZFJH202303. RSS is supported by the Laboratory Directed Research and Development (LDRD No. 23-051) of Brookhaven National Laboratory (BNL) and by RIKEN-BNL Research Center.

\bibliography{alpha_UV_ext}

\end{document}